\documentclass[12pt]{article}

\usepackage{scicite}
\usepackage{epsfig}
\usepackage{times}
\usepackage{amsmath}

\usepackage{amssymb}

\newcommand{\be}{\begin{equation}}
\newcommand{\ee}{\end{equation}}
\newcommand{\bea}{\begin{eqnarray}}
\newcommand{\eea}{\end{eqnarray}}

\renewcommand{\epsilon}{\varepsilon}

\newenvironment{sciabstract}{%
\begin{quote} \bf}
{\end{quote}}

\newcounter{lastnote}
\newenvironment{scilastnote}{%
\setcounter{lastnote}{\value{enumiv}}%
\addtocounter{lastnote}{+1}%
\begin{list}%
{\arabic{lastnote}.}
{\setlength{\leftmargin}{.22in}}
{\setlength{\labelsep}{.5em}}}
{\end{list}}

\title{Microwave-Induced Cooling of a Superconducting Qubit}

\author{Sergio O. Valenzuela,$^{1*}$ William D. Oliver,$^{2}$ David M. Berns,$^{3}$\\
Karl K. Berggren,$^{2\dagger}$ Leonid S. Levitov,$^{3}$ Terry P. Orlando$^{4}$\\
\\
 \normalsize{$^{1}$Massachusetts Institute of Technology (MIT) Francis Bitter Magnet Laboratory,}\\
  \normalsize{Cambridge, MA 02139, USA}\\
 \normalsize{$^{2}$MIT Lincoln Laboratory, 244 Wood Street, Lexington, MA 02420, USA}\\
 \normalsize{$^{3}$Department of Physics, MIT, Cambridge, MA 02139, USA}\\
 \normalsize{$^{4}$Department of Electrical Engineering and Computer Science, MIT,} \\
 \normalsize{ Cambridge, MA 02139, USA}\\
 \\
 \normalsize{$^\ast$To whom correspondence should be addressed; E-mail: sov@mit.edu } \\
 \normalsize{$^\dagger$Present address: MIT EECS Department}  }

\date{}

\begin{document}


\baselineskip24pt


\maketitle


\begin{sciabstract}
We demonstrated microwave-induced cooling in a
superconducting flux qubit. The thermal
population in the first-excited state of the qubit is driven to a
higher-excited state by way of a sideband transition.
Subsequent relaxation into the
ground state results in cooling. Effective temperatures as low as
$T_{\rm eff} \approx 3$ millikelvin are achieved for bath
temperatures $T_{\rm bath} = 30 - 400$ millikelvin, a cooling
factor between 10 and 100. This demonstration provides an analog
to optical cooling of trapped ions and atoms and
is generalizable to other solid-state quantum systems. Active
cooling of qubits, applied to quantum information science, provides a means for qubit-state preparation with improved fidelity and for suppressing decoherence in multi-qubit systems.
\end{sciabstract}

\newpage


Cooling dramatically affects the quantum dynamics of a system, suppressing dephasing and noise processes and
revealing an array of lower-energy quantum-coherent phenomena,
such as superfluidity, superconductivity, and the Josephson
effect.
Conventionally, the entire system under study is cooled with $^3$He-$^4$He cryogenic techniques. Although
this straightforward approach has advantages, such as cooling
ancillary electronics and providing thermal stability, it also has
drawbacks. In particular, limited cooling efficiency and poor heat
conduction at millikelvin temperatures limit the lowest
temperatures attainable.

\begin{figure}[t]
\vspace{-20mm}
\begin{center}\epsfig{file=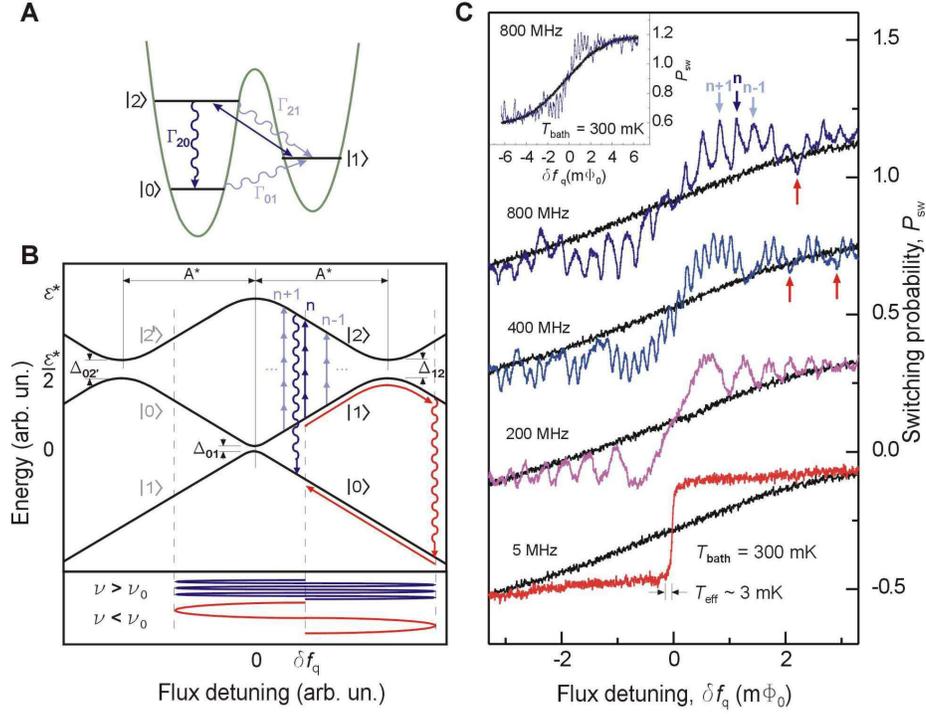,width=5in}\end{center}\vspace{-5mm}
\caption{\footnotesize{Sideband cooling in a flux qubit. (\textbf{A}) External excitation
transfers the thermal population from state $|1\rangle$ to state
$|2\rangle$ (straight line) from which it decays into the ground
state $|0\rangle$. Wavy lines represent spontaneous relaxation and
absorption, $\Gamma_{20}\gg \Gamma_{21},\Gamma_{01}$. The double
well is the flux-qubit potential comprising energy levels.
(\textbf{B}) Schematic band diagram illustrating the resonant and
adiabatic sideband cooling of the ac-driven qubit. $|1\rangle \to
|2\rangle$ transitions are resonant at high driving frequency
$\nu$ (blue lines) and occur via adiabatic passage at low $\nu$
(red lines). $\Delta_{01}$ and $\Delta_{12}$ are the tunnel splittings between
$|0\rangle$ - $|1\rangle$ and $|1\rangle$ - $|2\rangle$.
(\textbf{C}) Cooling induced by ac-pulses with
driving frequencies $\nu$ = 800, 400, 200 and 5 MHz. State
$|0\rangle$ population $P_{\rm sw}$ versus flux detuning $\delta
f_{\rm q}$ for the cooled qubit and for the qubit in thermal
equilibrium with the bath (black lines, $T_{\rm bath} = 300\,{\rm
mK}$). Measurements for $\nu=$
800, 200 and 5 MHz are displaced vertically for clarity. (Inset)
$P_{\rm sw}$ versus $\delta f_{\rm q}$ over a wider range of flux
detuning; $\nu=$ 800 MHz.}} \label{fig1}
\end{figure}

A fundamentally different approach to cooling has been developed
and implemented in quantum
optics\cite{Chu98,Cohen-Tannoudji98,Phillips98,Leibfried03}. The
key idea is that the degrees of freedom of interest may be cooled
individually, without relying on heat transfer among different
parts of the system. By such directed cooling processes, the
temperature of individual quantum states can be reduced by many
orders of magnitude with little effect on the temperature of
surrounding degrees of freedom. In one successful approach, called
sideband cooling \cite{Wineland78,Neuhauser78,Marzoli94,Monroe95}, the
unwanted thermal population of an excited state $|1\rangle$ is
eliminated by driving a resonant sideband transition to a higher excited
state $|2\rangle$, whose population quickly relaxes
into the ground state $|0\rangle$ (Fig.~\ref{fig1}A). The
two-level subsystem of interest, here $\{|0\rangle$, $|1\rangle
\}$, is efficiently cooled if the driving-induced population
transfer to state  $|0\rangle$ is faster than the thermal
repopulation of state $|1\rangle$. The sideband method, originally
used to cool vibrational states of trapped ions and atoms, allows
several interesting extensions
\cite{Chu98,Cohen-Tannoudji98,Phillips98,Leibfried03,Perrin98,Vuletic98,Kerman00,Morigi00}.
For example, the transition to an excited state can be achieved by
nonresonant processes, such as adiabatic passage\cite{Perrin98},
or by adiabatic evolution in an optical
potential\cite{Vuletic98,Kerman00,Morigi00}. Other approaches,
such as optical molasses and evaporative cooling, have been
developed to cool the translational degrees of freedom of atoms to
nanokelvin temperatures, establishing the basis for the modern
physics of cold atoms \cite{Wieman99}.

Superconducting qubits are mesoscopic artificial atoms
\cite{Clarke88} which exhibit quantum-coherent dynamics
\cite{Makhlin01a} and host a number of phenomena known to atomic
physics and quantum optics, including coherent quantum
superpositions of distinct macroscopic states
\cite{Friedman00a,Wal00a}, time-dependent Rabi oscillations
\cite{Nakamura99a,Nakamura01,Vion02a,Yu02a,Martinis02a,Chiorescu03a,Saito05},
coherent coupling to microwave cavity photons
\cite{Chiorescu04,Wallraff04,Johansson06} and St\"{u}ckelberg oscillations via Mach-Zehnder
interferometry \cite{Oliver05,Sillanpaa06,Berns06}. In a number of
these experiments, qubit state preparation by a dc pulse or by
thermalization with the bath was used. It is tempting,
however, to extend the ideas and benefits of optical cooling to
solid-state qubits, because they present a high degree of quantum
coherence, a relatively strong coupling to external fields, and
tunability, a combination rarely found in other fundamental
quantum systems.

We demonstrate a solid-state analog to optical cooling
utilizing a niobium persistent-current qubit \cite{Orlando99a}, a
superconducting loop interrupted by three Josephson junctions
\cite{onlinematerial}. When the qubit loop is threaded with a dc
magnetic flux $f_{\rm q} \approx \Phi_0/2$, where $\Phi_0 \equiv
h/2e$ is the flux quantum ($h$ is Planck's constant), the qubit's potential energy exhibits a
double-well profile (Fig.~\ref{fig1}A), which can be tilted by
adjusting the flux detuning, $\delta f_{\rm q} = f_{\rm q} -
\Phi_0/2$, away from zero. The lowest-energy states of each well
are the diabatic qubit states of interest, $|0\rangle$ and
$|1\rangle$, characterized by persistent currents $I_{\rm q}$ with
opposing circulation, whereas the higher-excited states in each
well, e.g., $|2\rangle$, are ancillary levels that
form the ``sideband transition'' with the qubit.
In contrast to conventional sideband cooling, which aims to cool an ``external'' harmonic oscillator (e.g., ion trap potential) with an ``internal'' qubit (e.g., two-level system in an ion), our demonstration aims to cool an ``internal'' qubit by using an ancillary ``internal'' oscillator-like state [supporting online material (SOM) Text].

When the qubit is in equilibrium with
its environment, some population is thermally excited from the
ground-state $|0\rangle$ to state $|1\rangle$ according to $p_1/p_0 = \exp[-(\varepsilon_1
- \varepsilon_0)/k_{\rm B} T_{\rm bath}]$, where $p_{0,1}$ are the
qubit populations for energy levels $\varepsilon_{0,1}$, $k_{\rm
B}$ is the Boltzmann constant, and $T_{\rm bath}$ is the bath
temperature. To cool the qubit subsystem below $T_{\rm bath}$, in
analogy to optical pumping and sideband cooling, a microwave magnetic flux of
amplitude $A$ and frequency $\nu$ targets the $|1\rangle \to
|2\rangle$ transition, driving the state-$|1\rangle$ thermal
population to state $|2\rangle$ from which it quickly relaxes to
the ground state $|0\rangle$. The hierarchy of relaxation and absorption rates
required for efficient cooling, $ \Gamma_{20}\gg
\Gamma_{21},\Gamma_{01}$, is achieved in our system owing to a
relatively weak tunneling between wells, which inhibits the
inter-well relaxation and absorption processes,
$|2\rangle\to|1\rangle$ and $|0\rangle\to|1\rangle$, compared with
the relatively strong intra-well relaxation process
$|2\rangle\to|0\rangle$. This three-level system
behavior is markedly different from the population saturation
observed in two-level systems.

The cooling procedure illustrated in Fig.~\ref{fig1}A is
generalized to the energy-band diagram shown schematically in Fig.
\ref{fig1}B. The diabatic-state energies,
\be\label{eq:bands}
 \epsilon_{1,0} = \pm I_{\rm q}
 \delta f_{\rm q}, \quad
 \epsilon_{2',2} = \epsilon^\ast \pm I_{\rm q} \delta f_{\rm q},
 \ee
are linear in the flux detuning $\delta f_{\rm q}$, with the
energy $\epsilon^\ast\approx 25\,{\rm GHz}$ and $I_{\rm
q}=1.44\,{\rm GHz/m}\Phi_0$ in our device, and
exhibit avoided crossings $\Delta_{01} \approx 12$ MHz and
$\Delta_{12} = \Delta_{02'} \approx 100$ MHz due to quantum
tunneling through the double-well barrier (Fig.~\ref{fig1}A). The diabatic levels exchange roles at each avoided crossing,
and the energy band is symmetric about $\delta f_{\rm q} = 0$
\cite{negative_delta_f}.

Under equilibrium conditions, the average level populations
exhibit a thermally-broadened ``qubit step" about $\delta f_{\rm
q} = 0$ , the location of the $|0\rangle$ - $|1\rangle$
avoided crossing. This is determined from the switching
probability $P_{\rm {sw}}$ of the measurement superconducting
quantum interference device (SQUID) magnetometer, which follows the
$|0\rangle$ state population \cite{onlinematerial},
\be\label{eq:Psw} P_{\rm {sw}}=\frac{1}{2} (1+Fm_0), \quad m_0  =\tanh\frac{\epsilon}{2k_B T},  \ee
where $F$ is the fidelity of the measurement, $m_0=
p_0-p_1$ is the
equilibrium magnetization that results from the qubit populations
$p_{0,1}$, $T = T_{\rm bath}$, and $\varepsilon = \varepsilon_1 -
\varepsilon_0 \propto \delta f_{\rm q}$ as inferred from Eq.
\ref{eq:bands}. In the presence of microwave excitation targeting
the $|1\rangle \to |2\rangle$ transition, the resultant
cooling, which we will later quantify in
terms of an effective temperature $T_{\rm eff} <
T_{\rm bath}$, acts to increase the ground-state population and,
thereby, sharpen the qubit step. This cooling signature is evident in Fig.~\ref{fig1}C, where we show
the qubit step before and after applying a cooling pulse at
several frequencies for $T_{\rm bath} = 300$ mK.

The cooling presented in Fig.~\ref{fig1}, B and C, exhibits a rich structure as a function of driving
frequency and detuning, resulting from the manner in which state $|2\rangle$ is accessed. The $|1\rangle \to |2\rangle$ transition
rate can be described by a product of a resonant factor and an
oscillatory Airy factor \cite{Berns06}. The former dominates at
high frequencies (800 and 400 MHz), where well-resolved resonances
of $n$-photon transitions are observed, as illustrated in Fig.
\ref{fig1}B (transition in blue) and Fig.~\ref{fig1}C (top traces
and inset). The cooling is thus maximized near the
detuning values matching $\epsilon_2-\epsilon_1=nh\nu$ (downward
arrows in Fig.~\ref{fig1}C). At intermediate frequencies (400 and
200 MHz), the Airy factor becomes more prominent and accounts for
the St\"{u}ckelberg-like oscillations that modulate the intensity
of the $n$-photon resonances \cite{Oliver05,Berns06}. Below $\nu=200\,{\rm MHz}$, although
individual resonances are no longer discernible, the modulation
envelope persists due to the coherence of the Landau-Zener
dynamics at the $\Delta_{12}$ avoided crossing \cite{Berns06}. The
$|1\rangle\to|2\rangle$ transition becomes weak near the zeros of
the modulation envelope, where we observe less efficient cooling,
or even slight heating (e.g., upward arrows in Fig.~\ref{fig1}C, 800 and 400 MHz). This is a result of the
$|0\rangle\to|1\rangle$ transition rate which, although relatively
small, $\Delta_{01}^2 \ll \Delta_{12}^2$, acts to excite the qubit
when the usually dominant $|1\rangle\to|2\rangle$
transition rate vanishes. At low frequencies [$\nu \lesssim
\nu_0=(\Delta_{12}^3/A^\ast)^{1/2} \approx 10\,{\rm MHz}$], the
state $|2\rangle$ is reached via adiabatic passage (Fig.~\ref{fig1}B, red lines) and the population transfer and cooling become
conveniently independent of detuning (see $\nu=5 ~ {\rm MHz}$ in
Fig.~\ref{fig1}C).

\begin{figure}\vspace{-20mm}
\begin{center}\epsfig{file=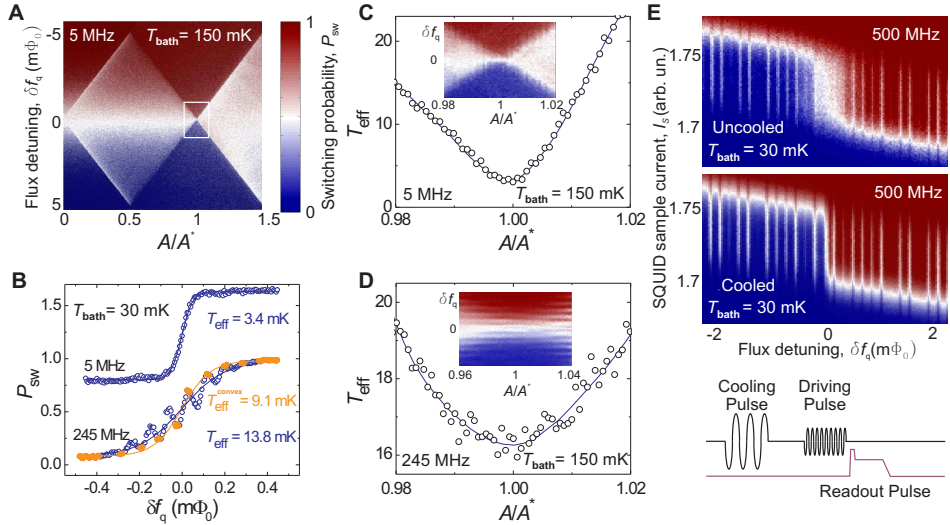,width=5in}\end{center}\vspace{-5mm}
\caption{\footnotesize{Optimal cooling parameters and effective temperature. (\textbf{A}) State
$|0\rangle$ population $P_{\rm sw}$ versus flux detuning, $-A^*/2
\lesssim \delta f_{\rm q} \lesssim A^*/2$, and driving amplitude
$A$ with $\nu$ = 5 MHz, $t_p = 3\,\mu{\rm s}$, and $T_{\rm bath}$
= 150 mK. Optimal conditions for cooling are realized at $A =
A^\ast$, where $A^*$ is defined in Fig.~1B. (\textbf{B}) Effective
temperature $T_{\rm{eff}}$. Qubit steps measured at $\nu=$ 5 and
245 MHz (circles) and best fits to Eq. \ref{eq:Psw}. At 245 MHz,
the aggregate temperature fitting (blue, $T_{\rm{eff}}$ = 13.8 mK) and the convex fitting (orange, $T_{\rm{eff}}$ = 9.1 mK) are shown. $T_{\rm bath}$ = 30 mK. (\textbf{C} and
\textbf{D}) (Inset) Detail of the region $A\approx A^*$ [white box
in (\textbf{A})] for $\nu$ = 5 MHz (top) and $\nu$ = 245 MHz
(bottom). In each case, $T_{\rm{eff}}$ is extracted from the qubit
step as in (\textbf{B}). Lines are guides for the eye; $t_{\rm p}
= 3~\mu$s, $T_{\rm bath}$ = 150 mK. (\textbf{E}) Spectroscopy of
uncooled (top) and cooled (middle) qubit (5 MHz, 3-$\mu$s cooling pulse) at
$T_{\rm bath}=30$ mK. Cumulative switching-probability
distribution as a function of $I_{\rm{s}}$ and $\delta f_{\rm{q}}$
under 500-MHz ac excitation.}} \label{fig2}
\end{figure}

Maximal cooling occurs near an optimal driving amplitude (Fig.~\ref{fig2}). Fig.~\ref{fig2}A
shows the $|0\rangle$ state population $P_{\rm sw}$ measured as a
function of the microwave amplitude $A$ and flux detuning $\delta
f_{\rm q}$ for frequency $\nu=5\,{\rm MHz}$. The adiabatic passage
regime, realized at this frequency, is particularly simple to
interpret, although higher frequencies allow for an analogous
interpretation. Cooling and the diamond feature of size
$A^* = \epsilon ^*/ 2I_{\rm q}$ can be understood in terms of the
energy band diagram (Fig.~\ref{fig1}B). For amplitudes $0 \leq A
\leq A^*/2$, population transfer between states $|0\rangle$ and
$|1\rangle$ occurs when $A>|\delta f_{\rm q}|$, such that the
sinusoidal flux reaches the $\Delta_{01}$ avoided crossing; this
defines the front side of the observed spectroscopy diamond
symmetric about the qubit step. For amplitudes $A^*/2 \leq A \leq
A^*$, the $\Delta_{12}$ ($\Delta_{02'}$) avoided crossing
dominates the dynamics, resulting in a second pair of thresholds
$A = A^*-|\delta f_{\rm q}|$, which define the back side of the
diamond. As the diamond narrows to the point $A = A^*$, cooling is observed. There only
one of the two side avoided crossings is reached and, thereby,
strong transitions with relaxation to the ground state result for
all $\delta f_{\rm q}$, yielding the sharpest qubit step. For $A > A^*$, both
side avoided crossings $\Delta_{12}$ and $\Delta_{02'}$ are reached
simultaneously when $|\delta f_{\rm q}|\lesssim A-A^\ast$, leading
again to a large population transfer between $|0\rangle$ and
$|1\rangle$.

When an ac field is applied, the qubit is no
longer in equilibrium with the bath, but it can still be
well-characterized by an effective temperature $T_{\rm{eff}}$
using Eq. \ref{eq:Psw} with $T=T_{\rm{eff}}$. This is illustrated in Fig.~\ref{fig2}B for $\nu=5$ MHz and
$\nu=245$ MHz ($T_{\rm{bath}}=30$ mK). At $\nu=5$ MHz, the qubit
step clearly follows Eq. \ref{eq:Psw}, as shown with a fit line
for $T_{\rm {eff}} =3.4$ mK. At 245 MHz, individual multiphoton
resonances are evident, and $P_{\rm{sw}}$ is a non-monotonic
function of $\delta f_{\rm{q}}$. In this case, $T_{\rm {eff}}$ is
still a useful parameter to quantify the effective cooling, but it
should be interpreted as an aggregate temperature over all
frustrations. Alternatively, because the cooling is maximized at
individual resonances, one may perform a convex fitting of Eq.
\ref{eq:Psw}, where only the solid (orange) symbols are taken into
account to determine the effective temperature at the resonance
detunings. The convex effective temperature $T_{\rm{eff}}^{\rm
convex} = 9.1$ mK is smaller than the aggregate value
$T_{\rm{eff}} = 13.8$ mK. In the remainder of the paper, we refer
to the more conservative effective temperature obtained using the
aggregate definition.

Figure~\ref{fig2}, C and D, show the variation of $T_{\rm
eff}$ about $A = A^*$ for $\nu=5$ MHz and $\nu=245$ MHz,
respectively, in the region marked with a white rectangle in Fig.
\ref{fig2}A (insets show the raw data). As seen in these figures,
$T_{\rm eff}$ typically presents a minimum, where the cooling is
most efficient and from which $A^*$ can be determined.

To determine whether the observation of a sharp
qubit step proves that the system makes transitions to the ground
state, as opposed to selectively populating an excited state with
the same magnetization, we measured the
excitation spectra of the ``pre-cooled" qubit and of the qubit in thermal equilibrium with the
bath (Fig.
\ref{fig2}E). In the former, a weak ac excitation was applied
immediately after the cooling pulse (time-lag less than 100 ns), well before the system equilibrates by warming up to the
bath temperature (see below). By comparing the excitation spectra
of the equilibrium and cooled systems (Fig.~\ref{fig2}E, $T_{\rm
bath} =30$ mK), we note that, although cooling markedly reduces
the step width, making the qubit much colder, the excitation
spectrum remains unchanged. Because the ac excitation is resonant
with the $|0 \rangle \to |1 \rangle$ transition only, this
strongly indicates that the population in a cooled qubit is in the
ground state.

\begin{figure}\vspace{-20mm}
\begin{center}\epsfig{file=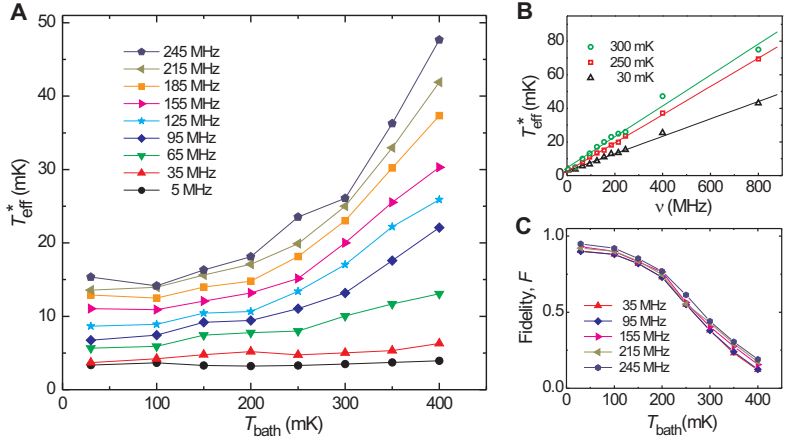,width=4.5in}\end{center}\vspace{-5mm}
\caption{\footnotesize{Effective temperature $T_{\rm eff}^*$ for $A=A^*$ and measurement
fidelity $F$. (\textbf{A}) $T_{\rm eff}^*$ versus $T_{\rm{bath}}$ at
the indicated driving frequencies $\nu$. $T_{\rm eff}^*$ increases
with $T_{\rm{bath}}$ at high $\nu$, but remains constant at low
$\nu$. (\textbf{B}) $T_{\rm eff}^*$ versus $\nu$ for different
$T_{\rm{bath}}$. Lines are linear fits.
(\textbf{C}) $F$ versus $T_{\rm{bath}}$ at the indicated $\nu$. A pulse width $t_{\rm p}=3 ~ \mu$s was
used in all cases.}} \label{fig3}
\end{figure}

Figure~\ref{fig3}, A and B, summarize the dependence of
$T_{\rm{eff}}^*=T_{\rm{eff}}(A^*)$ on the dilution refrigerator
temperature $T_{\rm{bath}} = 30 - 400$~mK for several frequencies
$\nu$, spanning the resonant sideband to the adiabatic passage
limits, with a fixed pulse width $t_{\rm p}= 3~ \mu$s. In Fig.
\ref{fig3}A, at large $\nu$, $T_{\rm{eff}}^*$ exhibits a monotonic
increase with $T_{\rm{bath}}$, which becomes less pronounced as
$\nu$ decreases. In the adiabatic passage limit, e.g. $\nu
=5$ MHz, $T_{\rm{eff}}^* \approx 3$ mK is practically constant and
reaches values that, notably, can be more than two
orders of magnitude smaller than $T_{\rm{bath}}$.
In Fig.~\ref{fig3}B, $T_{\rm eff}^*$ is observed to increase
linearly with $\nu$ for different values of $T_{\rm bath}$. Because
the number of resonances in the qubit-step region is inversely
proportional to $\nu$, the cooling at the individual resonances
depends only weakly on $\nu$ when using the convex definition $T_{\rm{eff}}^{\rm convex}(A^*)$.

Figure~\ref{fig3}C displays the measurement-fidelity $F$ versus
$T_{\rm{bath}}$. Although the qubit is effectively cooled, $T_{\rm
eff}^* \ll T_{\rm{bath}}$, over the range of $T_{\rm{bath}}$ in
Fig.~\ref{fig3}, A and B, the readout SQUID is not
actively cooled, and its switching current distribution broadens
with $T_{\rm{bath}}$ (fig.~S2). At high temperatures, the fidelity $F$, defined in Eq. \ref{eq:Psw},
becomes too small to discriminate the two qubit states; this is
independent of the qubit's effective temperature, which remains
$\sim$3 mK at all values of $T_{\rm{bath}}$. We observe that the fidelity
$F$ is larger than 0.8 for $T_{\rm{bath}} < 100$ mK, remains above
0.5 at $^3$He refrigerator temperatures, but drops to $F \approx
0.1$ at $T_{\rm{bath}}=400$~mK, limiting our ability to measure
the qubit state at higher temperatures (SOM Text).

\begin{figure}[t] \vspace{-20mm}
\begin{center} \epsfig{file=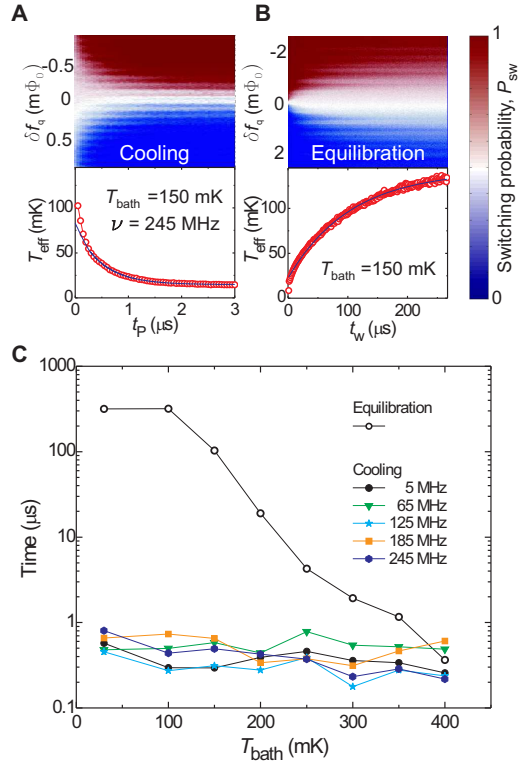,width=3in} \end{center}\vspace{-5mm}
\vspace{-5mm}\caption{\footnotesize{Dynamics of cooling and equilibration. (\textbf{A}) (Upper panel) State $|0
\rangle$ population $P_{\rm sw}$ as a function of $\delta f_{\rm
q}$ and cooling pulse width $t_{\rm p}$ ($\nu$ = 245 MHz). (Lower
panel) $T_{\rm{eff}}$ versus $t_{\rm p}$ extracted from upper panel
(circles) and exponential fit (blue line) with $\sim$ 1-$\mu$s
time constant. (\textbf{B}) (Upper panel) State $|0 \rangle$
population $P_{\rm sw}$ as a function of $\delta f_{\rm q}$ and
waiting time $t_{\rm w}$ after the cooling pulse ($t_{\rm p} =
3~\mu$s and $\nu =$ 5 MHz). (Lower panel) $T_{\rm{eff}}$ versus
$t_{\rm w}$ extracted from upper panel (circles) and exponential
fit (blue line) with $\sim$ 100-$\mu$s time constant.
$T_{\rm{bath}}$ = 150 mK. (\textbf{C}) Characteristic
equilibration and cooling times for different $T_{\rm{bath}}$.
Cooling is performed at the indicated frequencies.}} \label{fig4}
\end{figure}

The cooling and equilibration dynamics of the qubit are summarized
in Fig.~\ref{fig4} ($T_{\rm bath}=150$ mK). Cooling a qubit in equilibrium with the bath requires a characteristic cooling
time. In turn, a cooled qubit is effectively colder than its
environment, a non-equilibrium condition, which over a
characteristic equilibration time will thermalize to the
environmental bath temperature. This relation between cooling
and equilibration times determines the facility of cooling the
qubit and performing operations while still cold. Fig.~\ref{fig4}, A
and B, show the time evolution at cooling and
warming up of the qubit step. The top panels show $P_{\rm sw}$ as
a function of $\delta f_{\rm q}$ and cooling-pulse length $t_{\rm
p}$ (Fig.~\ref{fig4}A, $\nu = 245$ MHz), and as a function of
$\delta f_{\rm q}$ and waiting-time $t_{\rm w}$ after pre-cooling
with a 5 MHz pulse (Fig.~\ref{fig4}B) (for $t_{\rm p}$ and $t_{\rm
w}$ definition, see fig.~S1). Note the difference in the time
scales, where it is observed that substantial cooling is
accomplished within 1 $\mu$s (Fig.~\ref{fig4}A), but equilibration
occurs over a much longer time scale (Fig.~\ref{fig4}B). Fitting
to Eq. \ref{eq:Psw} yields $T_{\rm eff}$ as a function of $t_{\rm
p}$ and $t_{\rm w}$ (Fig.~\ref{fig4}, A and B, bottom panels). The
near exponential behavior of $T_{\rm eff}$ versus $t_{\rm p}$ and
$t_{\rm w}$ allows one to infer the characteristic cooling and
equilibration times as defined by an exponential fitting (solid
blue lines), which are summarized in Fig.~\ref{fig4}C. Notably,
the cooling characteristic time is nearly independent of both
$\nu$ and $T_{\rm bath}$ and, on average, is about 500 ns. In
contrast, at the base temperature of the dilution refrigerator,
the equilibration time is about three orders of magnitude longer, 300
$\mu$s, and remains one order of magnitude longer at 250 mK, a
temperature that is accessible with $^3$He refrigerators.

The minimum qubit effective temperature demonstrated in this work
was estimated to be $T_{\rm eff} \approx 3$~mK. This value is
consistent with the inhomogeneously broadened linewidth observed in the experiment, which likely places a
lower limit on the measurable minimum temperature. The
microwave-induced cooling presented here can be applied
to problems in quantum information science, including
ancilla-qubit reset for quantum error correcting codes and
quantum-state preparation, with implications for improved fidelity
and decoherence in multi-qubit systems. This approach, realized in
a superconducting qubit, is generalizable to other solid-state
qubits and can be used to cool other on-chip elements, e.g. the qubit circutry or resonators




\begin{thebibliography}{10}
\bibitem{onlinematerial}
Materials and methods are available as supporting material on
Science Online.
\end{thebibliography}

\bibliographystyle{Science}


\begin{scilastnote}
\item We thank A. J. Kerman, D. Kleppner, and A. V. Shytov for helpful discussions; and
V. Bolkhovsky, G. Fitch, D. Landers, E. Macedo, P. Murphy, R.
Slattery, and T. Weir at MIT Lincoln Laboratory for technical
assistance. This work was supported by Air Force Office of Scientific Research
(grant F49620-01-1-0457) under the DURINT program and partially by the Laboratory for Physical Sciences. The work at Lincoln
Laboratory was sponsored by the US Department of Defense under Air Force Contract
No. FA8721-05-C-0002.
\end{scilastnote}

\newpage
\section*{Materials and Methods}

\begin{figure}[h]
\vspace{40mm}
 \begin{center} \epsfig{file=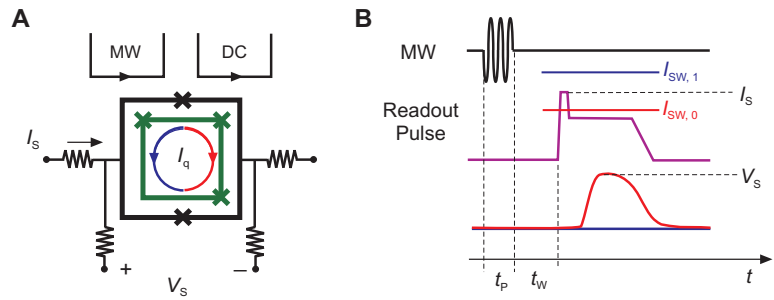,width=4in} \end{center}
\vspace{5mm}\caption{(Fig. S1)\footnotesize{ Schematic flux qubit and measurement scheme.
(\textbf{A}) Qubit (inner green loop) and inductively coupled DC
SQUID magnetometer (outer black loop). Josephson junctions are
indicated by crosses. Red and blue arrows indicate the two
directions of the persistent current in the qubit, $I_{\rm q}$. DC
and pulsed MW magnetic fields $f(t) = f^{\rm dc} + f^{\rm ac}(t)$
control the qubit and drive transitions between its quantum
states. (\textbf{B}) The qubit state is inferred by sampling the
SQUID with a pulsed current $I_{\mathrm{s}}$ and performing
voltage readout of the switching pulse, $V_{\mathrm{s}}$. The
current pulse $I_{\mathrm{s}}$ is set to maximize the difference
in the SQUID switching probability between the two qubit states
which couple to the SQUID through the associated persistent
currents, $I_{\rm q}$. The SQUID only switches when the qubit is
in state $|0\rangle$ and the switching probability $P_{\rm sw}$
measures its population.
}}
 \label{fig1S}
\end{figure}

\paragraph{Measurement Scheme}
The qubit consists of a superconducting loop interrupted by three
Josephson junctions (Fig.~S\ref{fig1S}A), one of which has a
reduced cross-sectional area. DC and pulsed microwave (MW)
magnetic fields generate a time-dependent magnetic flux $f(t) =
f^{\rm dc} + f^{\rm ac}(t)$ threading the qubit. Transitions
between the qubit states are driven by the pulsed MW flux $f^{\rm
ac}(t)= A\cos 2\pi \nu t$ of duration $t_{\rm p}=10\,{\rm
ns}-3\,\mu{\rm s}$ (Fig.~S\ref{fig1S}B) with frequency $\nu$, and
amplitude $A$, where $A$ is proportional to the MW-source voltage
$V_{\rm ac}$. The qubit states are read out using a DC-SQUID, a
sensitive magnetometer that distinguishes the flux generated by
the qubit persistent currents, $I_{\rm q}$. After a delay $t_{\rm
w}$ following the excitation, the readout is performed by driving
the SQUID with a 20-ns ``sample'' current $I_{\rm{s}}$ followed by
a 20-$\mu$s ``hold'' current (Fig.~S\ref{fig1S}B). The SQUID will
switch to its normal state voltage $V_{\rm{s}}$ if $I_{\rm{s}} >
I_{\rm{sw,0}}$ ($I_{\rm{s}}
> I_{\rm{sw,1}}$), when the qubit is in state $|0\rangle$ ($|1\rangle$).
By sweeping $I_{\rm{s}}$ and flux detuning, while monitoring the
presence of a SQUID voltage over many trials, we generate a
cumulative switching distribution function (see Fig.~2E and Fig.
S2 below). By following a flux-dependent sample current
$I_{\rm{sw,0}} < I_{\rm{s}} < I_{\rm{sw,1}}$ we obtain the
switching probability $P_{\rm{sw}}$ that characterizes the
population of state $|0\rangle$, and that reveals the ``qubit
step" shown in Fig.~1C.

The experiments were performed in a dilution refrigerator with a
12-mK base temperature. The device was magnetically shielded with
4 Cryoperm-10 cylinders and a superconducting enclosure. All
electrical leads were attenuated and/or filtered to minimize
noise.

\paragraph{Device Fabrication and Parameters}
The device (Fig.~S\ref{fig1S}A) was fabricated at MIT Lincoln
Laboratory on 150 mm wafers in a fully-planarized niobium trilayer
process with critical current density $J_{\rm{c}} \approx 160$
$\rm{A}/ \rm{cm}^2$. The qubit's characteristic Josephson and
charging energies are $E_{\rm{J}} \approx (2\pi\hbar)300$ GHz and
$E_{\rm{C}} \approx (2\pi\hbar)0.65$ GHz respectively, the ratio
of the qubit Josephson junction areas is $\alpha \approx 0.84$,
and the tunnel coupling $\Delta\approx (2\pi\hbar)0.01$ GHz. The
qubit loop area is 16 $\times$ 16 $\mu$m$^2$, and its self
inductance is $L_{\rm{q}} \approx 30$ pH. The SQUID Josephson
junctions each have critical current $I_{\rm{c0}} \approx 2$
$\mu$A. The SQUID loop area is 20 $\times$ 20 $\mu$m$^2$, and its
self inductance is $L_{\rm{S}} \approx 30$ pH. The mutual coupling
between the qubit and the SQUID is $M \approx 25$ pH.

\section*{Supporting Text}
\paragraph{Microwave cooling and optical resolved sideband cooling}
While the microwave cooling (MC) demonstrated in this work has
similarities with optical resolved sideband cooling (RSC), the
analogy is not a tautology. We 
describe here the similarities
and 
distinctions between the two cooling techniques.

In both cases, 
the spectra of the
energy levels can be reduced to the three-level system defined in
Fig.~1A (main text). Although, as described below, the origin of
the three levels is different for MC and RSC, cooling is similarly
achieved by driving the thermal population in state $|1
\rangle$ to an ancillary state $|2 \rangle$, from which it quickly
relaxes to state $|0 \rangle$.

In the case of RSC of an atom, the three-level system illustrated
in Fig.~1A (without the double-well potential) results from a
two-level atomic system (TLS) combined with a simple harmonic
oscillator (SHO) from the trap potential. Using the
notation $|\text{TLS state},\text{SHO state} \rangle$ with TLS
states $ \{|g\rangle,|e\rangle \}$ and SHO states $\{|n\rangle \}$
with $n=0,1,2, \ldots$, one can identify from Fig.~1A the
following: $|0\rangle_{\text{RSC}} \equiv |g,n \rangle$,
$|1\rangle_{\text{RSC}} \equiv |g,n+1 \rangle$, and
$|2\rangle_{\text{RSC}} \equiv |e,n \rangle$. Note that the TLS
represents an ``internal'' atomic state, whereas the SHO is an
``external'' trap state.

In the case of MC of a flux qubit, the double-well potential
illustrated in Fig.~1A comprises two coupled SHO-like wells.
The left and right wells correspond to the diabatic
states of the qubit, the clockwise and counterclockwise
circulating currents, and together form the qubit TLS.
Each well independently has a series of SHO-like states. Using the
same notation $|\text{TLS state},\text{SHO state} \rangle$ with
TLS states $ \{|g\rangle,|e\rangle \}$, associated respectively to the
left and right well in Fig.~1A, one can identify the following:
$|0\rangle_{\text{MC}} \equiv |g,0 \rangle$,
$|1\rangle_{\text{MC}} \equiv |e,0 \rangle$, and
$|2\rangle_{\text{MC}} \equiv |g,1 \rangle$. Further
higher-excited states are not explicitly shown in Fig.~1A.
Note that all states here are ``internal,'' and that the
SHO-nature of the left and right wells is limited by the degree of
tunnel coupling between wells.

From the above discussion, it is clear that in both RSC and MC
there is a TLS combined with a SHO, however the roles of the
levels interchange.  Considering an associated frequency
$\omega_{\text{TLS}} = E_{\text{TLS}} / \hbar $ for the TLS and a
plasma frequency $\omega_{\text{SHO}} = E_{\text{SHO}} / \hbar$,
the two cases can be summarized as follows:

\begin{itemize}
  \item In the RSC case, $\omega_{\text{SHO}} < \omega_{\text{TLS}}$, and it is the SHO that is cooled.
       One drives
       population from the TLS ground state with higher SHO energy,
       $|1\rangle_{\text{RSC}} \equiv |g,n+1 \rangle$, to the TLS
       excited-state with lower SHO energy, $|2\rangle_{\text{RSC}} \equiv |e,n
       \rangle$, from which it relaxes to the ground state $|0\rangle_{\text{RSC}}
       \equiv |g,n \rangle$.

       \item In the MC case, $\omega_{\text{TLS}} < \omega_{\text{SHO}}$, and it is the TLS that is cooled.
       One drives
       population from the TLS excited state with low SHO energy,
       $|1\rangle_{\text{MC}} \equiv |e,0 \rangle$, to the TLS
       ground-state with higher SHO energy, $|2\rangle_{\text{RSC}} \equiv |g,1
       \rangle$, from which it relaxes to the ground state $|0\rangle_{\text{MC}}
       \equiv |g,0 \rangle$.
\end{itemize}

Thus, if one considers a three-level state configuration without
tagging the states with the terms ``external" or ``internal",
in one cooling cycle, one cools the subsystem of interest
$\{|0\rangle, |1\rangle \}$ by driving transitions to an ancillary
state $|2 \rangle$, which relaxes quickly to the ground state. For
MC, the cooled subsystem is the TLS, whereas for RSC, the cooled
subsystem is the SHO. Note that in the RSC case, because it is the
SHO that is cooled, the cooling cycles can be cascaded to cool the
multiple SHO states $n=0,1,2,\ldots$. However, in the MC case, it
is the TLS that is cooled, which requires only a single
cooling cycle from the TLS excited state to its ground state. In
the MC case described in this work, all transitions are allowed,
and the energy levels are widely tunable. However, in many
cases of RSC of atoms, certain transitions may be forbidden, and there
is typically only limited energy-band tunability.

\begin{figure}
 \begin{center} \epsfig{file=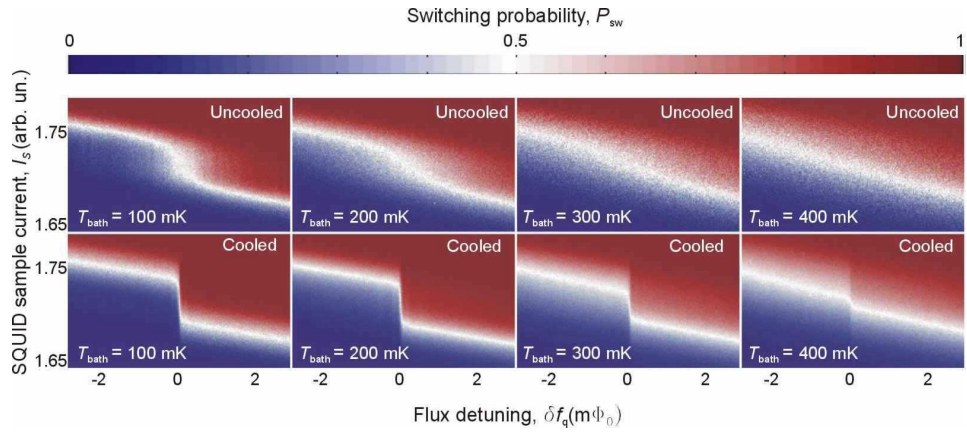,width=5in} \end{center}
 \caption{(Fig. S2)\footnotesize{ Cumulative switching distribution of the qubit as a function of $I_{\rm{s}}$ and $\delta
    f_{\rm{q}}$ of the qubit in equilibrium with the bath (top) and of
    the cooled qubit (bottom) at different $T_{\rm bath}$. The cooling
    pulse $t_{\rm p} = 3~\mu$s, $\nu$ = 5 MHz. Although the readout SQUID switching distribution
    broadens as $T_{\rm bath}$ increases, reducing the readout fidelity, the cooled qubit step remains sharp $(T_{\rm eff} \sim
    3$~mK).}}
 \label{fig2S}
\end{figure}

\paragraph{Effective temperature and fidelity vs. bath temperature}
The qubit step and the SQUID switching-current distribution
broaden with $T_{\rm{bath}}$. The qubit can be cooled effectively,
$T_{\rm eff}^* \ll T_{\rm{bath}}$, over the range of
$T_{\rm{bath}}$ in Fig.~3A and Fig.~3B. However, the readout SQUID
is not actively cooled, and its switching current distribution
broadens with $T_{\rm{bath}}$. This is observed in Fig.
S\ref{fig2S}, where we plot the cumulative switching-distribution
as a function of $I_{\rm s}$ and $\delta f_{\rm q}$ of uncooled
and cooled qubit (5 MHz, 3-$\mu$s cooling pulse) at different
$T_{\rm bath}$. At high temperatures, the switching-current
distribution becomes broad and the measurement fidelity $F$ drops
to values that become too small to discriminate the two qubit
states; this is independent of the qubit's effective temperature,
which remains about 3 mK at all $T_{\rm{bath}}$.

\clearpage
\section*{Supporting Figures}


\end{document}